\documentclass[preprint,12pt]{aastex}
\usepackage{emulateapj5}
\usepackage{apjfonts}
\usepackage{epsf}
\usepackage{graphicx}
\makeatletter

\newenvironment{inlinefigure}{%
\def\@captype{figure}%
\noindent\begin{minipage}{0.999\linewidth}\begin{center}}
{\end{center}\end{minipage}\smallskip}
\makeatother

\lefthead{INOUE and NAGASHIMA}
\righthead{Analytic Approach to the Cloud-in-cloud Problem for
Non-Gaussian Density Fluctuations}
\newcommand{\del}{\partial}

\newcommand{\x}{{\mathbf{x}}}

\newcommand{\z}{{\mathbf{z}}}

\newcommand{\K}{{\mathbf{k}}}

\newcommand{\ti}{\textit}
\newcommand{\f}{\frac}

\newcommand{\bb}{\bibitem}
\newcommand{\BF}{\begin{inlinefigure}\begin{center}}
\newcommand{\EF}{\end{center}\end{inlinefigure}}
\newcommand{\BE}{\begin{equation}}
\newcommand{\EE}{\end{equation}}
\newcommand{\BEA}{\begin{eqnarray}}
\newcommand{\EEA}{\end{eqnarray}}
\begin{document}

\submitted{\rm\it NAOJ-Th-Ap2001, No.57}

\title{Analytic Approach to the Cloud-in-cloud Problem for
Non-Gaussian Density Fluctuations}		
\author{Kaiki Taro Inoue and Masahiro Nagashima}
\affil{Division of Theoretical Astrophysics,
National Astronomical Observatory,
2-21-1 Osawa, Mitaka, Tokyo 181-8588, Japan;
\email{tinoue@th.nao.ac.jp, masa@th.nao.ac.jp}}

\begin{abstract}
We revisit the cloud-in-cloud problem for non-Gaussian density
fluctuations. We show that the extended Press-Schechter (EPS) formalism for
non-Gaussian fluctuations has a flaw in describing mass functions
regardless of type of filtering.  As an example, we consider
non-Gaussian models in which density fluctuations at a point obeys a
$\chi^2$ distribution with $\nu$ degrees of freedom.  We find that mass
functions predicted by using an integral formula proposed by Jedamzik, and
Yano, Nagashima \& Gouda, properly taking into account correlation
between objects at different scales, deviate from those predicted by
using the EPS formalism, especially for strongly non-Gaussian
fluctuations. Our results for the mass function at large mass scales 
are consistent with those by Avelino \& Viana 
obtained from numerical simulations.
\end{abstract}

\section{Introduction}
How many objects with mass $M$ are there in our universe?  This question
has been one of main interests in the field of cosmological structure
formation. Formation process of cosmological objects such as galaxy
clusters and galaxies is well understood qualitatively in the context of
the hierarchical clustering scenario based on a cold dark matter (CDM)
model.  In order to compute the number density of collapsed objects or
\ti{mass function}, one must deal with gravitational non-linear
growth of small density perturbations. The most direct way is to perform
$N$-body simulations. However, performing simulations for a large number
of models on wide range of scales is a very difficult task because of
the limit of the computation time and available amount of memory.
Therefore, it is of great importance to derive analytic formulae that
accurately describe the result of $N$-body simulations. Among them, the
Press-Schechter formula (Press \& Schechter 1974; hereafter PS) has been
a most successful one and applied to a wide class of structure formation
models.

Nevertheless, the PS formalism has a flaw in describing the number of
collapsed objects.  Because the underdense region of smoothed density fluctuations
is not taken into account in the formalism, integration of the mass
function over whole range of mass does not yield a unity.  Even if the
density fluctuation smoothed on mass scale $M$ is less than a critical
density fluctuation $\delta_c$, there is a chance that the density
fluctuation smoothed on larger mass scale $M'>M$ is larger than
$\delta_c$. This is called the ``cloud-in-cloud'' problem. Press \&
Schechter (1974) simply multiplied the mass function by the ``PS fudge
factor of two''in the case of Gaussian random fields.

The cloud-in-cloud problem for Gaussian random fields have been
partially solved by Peacock \& Heavens (1990) and Bond et al (1991)
using the so-called excursion set formalism and by Jedamzik (1995) and
Yano, Nagashima \& Gouda (1996) using an integral equation.  Consider a
density fluctuation $\delta_M$ smoothed on mass scale $M$ at a given
point.  Then one can regard a sequence of density fluctuations
$\delta_{M_1}, \delta_{M_2},\cdots $ in descending order of $M$ as a
trajectory of a ``particle.''  For fluctuations smoothed by the sharp
$k$-space filter, each trajectory is described by a Markovian
random walk of $\delta$ as a function of ``time'' $M$. Then we
analytically obtain the PS fudge factor of two.  However, for fluctuations smoothed by other
filters, no analytic result is known, since the correlation between
fluctuations with different scale renders the motion of a particle
non-Markov process.

In contrast, the cloud-in-cloud problem for non-Gaussian density
fluctuations had not been explored until recently.  As is well known, a
number of theoretical ``unstandard'' models including cosmic string
models, texture models, multiple fields models, and so on predict that
the primordial fluctuations are not Gaussian. Because the number of
collapsed dark matter halos at early epoch, \ti{e.g.} clusters at $z
\sim 1$ or galaxies at $z \gtrsim 5$, depends sensitively on the tails
of distribution function of initial density fluctuations, even a small
deviation from Gaussianity would cause a noticeable change in the
statistical property of those high-redshift objects.  To make
predictions on the number count of these rare objects, it is of crucial
importance to investigate how the PS formalism is extended to models
with non-Gaussian initial conditions.

Recently, there has been some progress on this issue based on the
so-called extended Press-Schechter (EPS) formalism in which the PS fudge
factor is again assumed to be a constant.  For Gaussian fluctuations
smoothed by the sharp $k$-space filter, the assumption is correct
because of the nature of the Markovian random walk of $\delta$. However,
in other cases, it is not known whether such assumption is correct,
especially for non-Gaussian models.  In the context of formalism
developed by Jedamzik (1995) and Yano, Nagashima \& Gouda (1996), the PS
fudge factor is equivalent to the inverse of conditional probability of
finding a region where $\delta_{M_1}\ge \delta_c$ of mass scale $M_1$
provided that it is totally included inside an isolated region where
$\delta_{M_2}=\delta_c$ of mass scale $M_2$.  In other words, the PS
fudge factor is not a constant and depends on smoothing scales $M_1,M_2$
in general.  This behavior was also noticed by Nagashima \& Gouda (1997)
using Monte Carlo simulations.  Although it has been claimed that the
EPS formalism provides a good fit to the mass function obtained from
$N$-body simulations for some non-Gaussian models (Robinson \& Baker
2000), one cannot immediately give a justification for the result. In
fact, recent numerical simulations of linear density fields showed that
the EPS formalism does not provide a good fit to the mass function for
strongly non-Gaussian probability distribution functions (PDFs) with
small variance $\sigma \lesssim 0.5$ which correspond to objects with
large mass, such as galaxy clusters at present (Avelino \& Viana
2000). For such small variances, the mass function is essentially
determined by the abundance of rare density peaks which are sensitive to
the non-Gaussianity of initial fluctuations. It is of crucial importance
to understand the role of the conditional probability for strongly
non-Gaussian density fluctuations, especially at large scales.

In this paper, we study non-Gaussian models in which one-point PDF of density
fluctuations smoothed on mass scale $M$ obeys a $\chi^2$ distribution
with $\nu$ degrees of freedom, which are very simple and widely used as
toy models in the literature (\ti{e.g.} Barreiro, Sanz, Mart\'\i
nez-Gonz\'alez \& Silk 1998). 
The degree of non-Gaussianity is
characterized by $\nu$. A $\chi^2_\nu$ distribution is strongly
non-Gaussian for small $\nu$ and converges to a Gaussian distribution in
the limit $\nu \rightarrow \infty$. 

In \S 2, we introduce a formalism for computing the mass
function. In \S 3, we describe simple models for which a density
fluctuation at a point obeys a $\chi^2_\nu$ distribution.  In \S 4, we
explore the property of the conditional probability and compare the
obtained mass functions with those predicted by using the EPS formalism.
In \S 5, the effect of mode correlation is discussed. 
In the last section, we summarize our result and draw our conclusions.

\section{Theory of Mass Function}
To compute mass functions analytically, PS made following assumptions:
(1) the overdense region collapses to an virialized object with mass $M$
when the linear density fluctuation $\delta$ smoothed on mass scale $M$
reaches a critical value $\delta_c$ which is a function of cosmic time;
(2) Each overdense region is independent and described by a
spherically symmetric collapse model which specifies $\delta_c$ (Tomita
1969; Gunn \& Gott 1972). Then the volume fraction of the collapsing
region at initial time with mass scale equal to or larger than $M$ is
simply given by
\begin{equation}
\psi (\ge\delta_c;M)=\int_{\delta_c}^\infty p(\delta;\sigma(M)) d\delta,
\end{equation}
where $p(\delta;\sigma(M))$ denotes a one-point PDF with variance
$\sigma^2(M)\equiv\langle\delta_{M}^{2}\rangle$ of the initial density
perturbation.  Now consider regions where the density fluctuation
$\delta$ smoothed on mass scale $M_1$ exceed $\delta_c$. Each region
should be totally contained inside an isolated collapsed object with
mass $M_2 \ge M_1$. Then we have
\begin{equation}
  \psi(\geq\delta_c;M_{1})=\int_{M_{1}}^{\infty}
   \frac{M_{2}}   
  {\bar\rho}n(M_{2})P(M_1 \vert M_2) dM_2,
\label{eq:jed0}
\end{equation} 
where $\bar \rho$ denotes the mean cosmic density, $P(M_1 \vert M_2)$ is
the conditional probability of finding a region $V_1$ of mass scale
$M_1$ where $\delta_1\equiv \delta(M_1)\ge \delta_c$ provided that $V_1$
is totally contained in an isolated overdense region $V_2$ where
$\delta_2 \equiv \delta(M_2)=\delta_c$ (Jedamzik 1995). 
In this formulation, one can
calculate the mass function by solving the integral equation
(\ref{eq:jed0}), once the conditional probability $P(M_{1}|M_{2})$ 
and the one-point PDF of the smoothed density fluctuations are given.
Note that the excursion set formalism developed by Bond et al. (1991) 
is essentially equivalent to this formalism. 
Because $V_2$ is an isolated region, 
the conditional probability is given by the
probability of the \ti{first upcrossing} of $\delta_2$ at the threshold
$\delta_c$ when smoothed on decreasing mass scales
\begin{equation}
P(M_1 \vert M_2)=p(\delta_1 \ge \delta_c \vert \delta_2=\delta_c, 
\delta(M) < \delta_c, ^\forall M>M_2), \label{eq:P12}
\end{equation}
assuming that the spatial correlation in the density fluctuations is
negligible.  Let us call a sequence of density fluctuations
$\delta(M_1), \delta(M_2), \cdots$ in descending order of $M$ as a
trajectory of a particle. For fluctuations $\delta(M)$ smoothed by the
sharp $k$-space filter whose phases of Fourier modes $\delta_\K$ are
uncorrelated, the motion of a particle is described by a Markovian
random walk. In this case, the conditional probability $P(M_1 \vert
M_2)$ does not depend on the state of a particle \ti{before} the
crossing $\delta_2=\delta_c$,
\begin{equation}
P(M_1 \vert M_2)=p(\delta_1 \ge \delta_c \vert \delta_2=\delta_c).
\label{eq:markov}
\end{equation}
In what follows we assume that equation (\ref{eq:markov}) gives a good
approximation of the conditional probability for fluctuations smoothed
by other filters (Yano, Nagashima \& Gouda 1996).  Then 
we only need to specify a one-point PDF and a two-point PDF of the smoothed density
fluctuations at a given point.  

If $P(M_{1}|M_{2})\equiv f$ does not depend on mass scales
$M_1$ and $M_2$, then the mass function is described by  
a formula similar to the PS formula, where the PS fudge-factor,
$f$, is related to the conditional probability as
\begin{equation}
f=[p(\delta_1 \ge \delta_c \vert \delta_2=\delta_c)]^{-1}=const. \label{eq:f}
\end{equation}
Let us first consider the Gaussian models.  The bivariate Gaussian
two-point PDF with vanishing means is given by
\begin{eqnarray}
\lefteqn{p_G(\delta_1,\delta_2;\sigma_1,\sigma_2,\tau)d\delta_{1}d\delta_{2}
=}\nonumber\\
&&\f{1}{2 \pi \sigma_1 \sigma_2 \sqrt{1-\tau^2}}
 \exp \Biggl [
- \f{ (\f{\delta_1}{\sigma_1})^2 + 
(\f{\delta_2}{\sigma_2})^2-2 \tau (\f{\delta_1}{\sigma_1} 
\f{\delta_2}{\sigma_2})} {2 (1-\tau^2)} \Biggr ]d\delta_{1}d\delta_{2},
\end{eqnarray}
where $\sigma_i^2$ denotes the variance of $\delta_i$, $\langle
\delta_i^2 \rangle$, and $\tau$ denotes the correlation coefficient,
$\langle\delta_{1}\delta_{2}\rangle/\sigma_{1}\sigma_{2}$.  Note that
$p_G$ is scale-invariant, \ti{i.e.}
$p_G(x_1,x_2;\sigma_1,\sigma_2,\tau)=\sigma_1^{-1} \sigma_2^{-1}
p_G(x_1/\sigma_1,x_2/\sigma_2,;1,1,\tau)$. 
For fluctuations smoothed by the sharp
$k$-space filter, $\tau=\sigma_2/\sigma_1$ (see appendix). Then the
two-point conditional PDF is written in terms of the one-point PDF as
$p_G(\delta_1 \vert \delta_2)=p_G(\delta_1,\delta_2)/p_G(\delta_2)=
p_G(\delta_1-\delta_2;\sqrt{\sigma_1^2-\sigma_2^2})$. Because the
Gaussian one-point PDF is also scale-invariant, we recover the constant
PS fudge factor, $f_G^{-1}=p_G(\delta_1\ge \delta_c \vert
\delta_2=\delta_c)=({2 \pi})^{-1/2} \int_0^\infty \exp [-\delta^2/2 ]
d\delta=1/2$. From equations (\ref{eq:jed0}) and
(\ref{eq:markov}), we obtain an explicit form of mass function,
\begin{equation}
n(M;\delta_c)dM=- \f{f \bar \rho}{M} \f{\del \psi(\ge \delta_c;M)}{\del M}dM.
\label{eq:n}
\end{equation}
In general, however, the PS fudge factor $f$ is not a constant and the
mass function $n(M)$ cannot be written explicitly as in equation
(\ref{eq:n}).  For instance, for Gaussian fluctuations smoothed by other
window functions, $P(M_{1}|M_{2})$ depends on smoothing scales $M_1$
 and $M_2$ (Nagashima 2001). 

In the non-Gaussian models, the mass scale dependence of $P(M_{1}|M_{2})$
must be always taken into account.  However, Koyama, Soda \& Taruya (1999;
hereafter KST) claimed that for generic non-Gaussian models, the
relation $p(\delta_1 \vert \delta_2)\simeq p(\delta_1-\delta_2)$ holds
for $M_1 \ll M_2$.  Consequently, we have
\begin{equation}
P(M_{1}|M_{2})\approx \int_{0}^\infty p(\delta_1,\sigma_1) d \delta_1. \label{eq:EPS}
\end{equation}
If the one-point PDF is scale-invariant, the above integration yields a constant
value.  Although obtaining a similar
relation for all scales $M_1 \le M_2$ is a more complex issue, KST
and some authors evaluated mass functions by solving equation (\ref{eq:n})
assuming a constant $f$ for various non-Gaussian models (Koyama, Soda \& Taruya 1999;
Robinson \& Baker 2000). From now
on, we call the PS formalism using the approximation described by
equations (\ref{eq:n}) and (\ref{eq:EPS}) in evaluating the mass
function, the extended PS (EPS) formalism.

Now let us evaluate the validity of the EPS formalism.  In the limit of
vanishing correlation coefficient, $\tau \rightarrow 0$, or equivalently $M_1 \ll
M_2$, the two-point PDF is written as a direct product of
one-point PDFs, $p(\delta_1,\delta_2)=p(\delta_1)p(\delta_2)$, which
gives
\begin{equation}
P(M_{1}|M_{2})=\int_{\delta_c}^\infty p(\delta_1; \sigma_1) 
d \delta_1. \label{eq:finv}
\end{equation}
On small mass scales with large variance, \ti{i.e.} $\sigma_1 \gg
\delta_c$, the lower limit of integration variable $\delta_c$ can be set
to zero as in the Gaussian fluctuations smoothed by the sharp $k$-space
filter, leading to a constant $f$. Hence, the EPS
approximation is valid for scales $M_1 \ll M_2$ and
$\sigma_1 \gg \delta_c$.  However, on large mass scales, where
$\sigma_{1}\lesssim\delta_{c}$, such approximation cannot be verified 
except for the Gaussian cases, since the contribution of integration
of $p(\delta_1; \sigma_1)$ in $\delta_1$ from $0$ to $\delta_c$ cannot
be negligible.  Therefore, the EPS approximation is not valid for
$\sigma_1 \lesssim \delta_c$ though $M_1 \ll M_2$.  This contradicts
the validity of the EPS approximation for $M_1 \ll M_2$ 
that KST have claimed. 
In generic non-Gaussian PDFs, the relation $p(\delta_1 \vert
\delta_2)\sim p(\delta_1-\delta_2)$ for $M_1 \ll M_2$ does not hold. In
fact, the class of PDFs of density fluctuations that satisfy $p(\delta_1
\vert \delta_2)\sim p(\delta_1-\delta_2)$ for $M_1 \ll M_2$ is very
limited\footnote{Although KST claimed that for $M_1 \ll M_2$, the
relation $p(\delta_1 \vert \delta_2)\sim p(\delta_1-\delta_2)$ holds for
generic non-Gaussian models, the statement is incorrect.  The statement
is correct if the cumulants satisfy a relation $ \langle \delta_1^m
\delta_2^{n-m} \rangle_c=\langle \delta_2^n \rangle_c$ for all $n>m$
rather than moments.  For Gaussian fluctuations, all the cumulants
$\langle \delta_1^m \delta_2^{n-m} \rangle_c$ vanish except for $n=2$
(the means are assumed to be zero).  However, for generic non-Gaussian
fluctuations, the cumulants do not necessarily vanish.  A condition for
moments that has been used in KST as an ingredient of the proof $
\langle \delta_1^m \delta_2^{n-m} \rangle=\langle \delta_2^n \rangle$ is
also incorrect for $n>2$. For instance, $\langle \delta_1^2 \delta_2^2
\rangle=2 \sigma_1^2 \sigma_2^2$ for $\tau=0$.}.  The Gaussian
fluctuations smoothed by the sharp $k$-space filter belong to this very
limited class.

It would be worthwhile to comment on relationship between the
conditional probability and the excursion set formalism.  If the
conditional probability satisfies $p(\delta_{1}|\delta_{2})=
p(\delta_{1}-\delta_{2})$, then the master equation can be
reduced to the diffusion equation by using the Kramers-Moyal expansion
which is used in the excursion set formalism.  However, it is clear that general 
PDFs do not necessarily satisfy the diffusion equation.  We need
to derive a proper two-point PDF that corresponds to the density fluctuation
distribution function under consideration.

\section{$\chi^2$ models}

We consider toy models in which the density fluctuation
$\delta_M$ smoothed on scale $M$ at a given point obey a $\chi^2_\nu$
PDF and we assume that the Fourier modes of 
fluctuations $\delta_\K$ are totally uncorrelated. The validity of this
assumption is discussed in \S5. 

Let $\x_i=(x_{1i},x_{2i}, \cdots, x_{Ni}),i=1,\cdots, \nu$ be independent
Gaussian $N$-dimensional vector variables.  We call the distribution of
$\z\equiv(\sum_{i=1}^\nu x_{1i}^2,\sum_{i=1}^\nu x_{2i}^2, \cdots,
\sum_{i=1}^\nu x_{Ni}^2) $ a $N$-point $\chi^2$ distribution with $\nu$
degrees of freedom. The one-point $\chi^2_\nu$ distribution is
\begin{equation}
p_{\chi^2_\nu}(x;\sigma)dx=\f{\sqrt{\nu/2}}{\sigma\Gamma(\nu/2)}
\left(\sqrt{\f{\nu}{2}}\f{x}{\sigma}\right)^{(\nu-2)/2}\exp
\left(-\sqrt{\f{\nu}{2}}\f{x}{\sigma}\right)dx,
\end{equation}
where $\Gamma(x)$ denotes the $\Gamma$ function.

Next, we derive the two-point PDF $p(z_1,z_2)$ of $\chi^2_\nu$
distribution with variance $\sigma_1$, $\sigma_2$ and correlation
coefficient $\tau$. Let us consider variables $x_1,x_2$ which obey a
two-point Gaussian distribution $p_G(x_1,x_2;s_1,s_2,\epsilon)$ with
vanishing means. By changing the variables as $z_1=x_1^2$, $z_2=x_2^2$,
one obtains
\begin{eqnarray}
p_{\chi_1^2}(z_1,z_2;\tau)dz_{1}dz_{2}
&=&
\f{1}{2 \pi s_1 s_2 \sqrt{(z_1 z_2)(1-\epsilon^2)}}
\cosh \Biggl[ \f{\epsilon\sqrt{z_1 z_2}}{s_1 s_2 (1-\epsilon^2)}\Biggr]\nonumber\\
&&\times
 \exp \Biggl [-\f{s_2 z_1/s_1+
s_1 z_2/ s_2}{2 s_1 s_2 (1-\epsilon^2)} \Biggr ]dz_{1}dz_{2},
\end{eqnarray}
where $s_i^2\equiv \langle x_i^2 \rangle =\langle z_i \rangle $,
$\sigma_i^2=2 s_i^4$ and $\tau=(\langle z_1 z_2 \rangle - \langle z_1
\rangle \langle z_2 \rangle )/\sigma_1 \sigma_2 =\epsilon^2$.  The
corresponding characteristic function is
\begin{equation}
\phi_{\chi^2_1}(t_1,t_2)=[1-2 i s_1 s_2 (s_1 t_1
s_2^{-1}+s_2 t_2
s_1^{-1})-4 s_1^2 s_2^2 (1-\epsilon^2)t_1 t_2]^{-1/2}. \label{eq:cha1} 
\end{equation}
The characteristic function for
$p_{\chi^2_\nu}(z_1,z_2;\sigma_1,\sigma_2,\tau)$ is given by
$\phi_{\chi^2_\nu}=(\phi_{\chi^2_1})^\nu$ where $\langle z_i \rangle=\nu
s_i^2$, Var$(z_i)= \sigma^2_i=2 \nu s_i^4$ and $\tau=\epsilon^2$, since
each variable $z_i$ is written as a sum of independent random
variables. From two-dimensional Fourier transform of
$\phi_{\chi^2_\nu}(t_1,t_2)$, we finally obtain the two-point PDF of
$\chi^2_\nu$ distribution,
\begin{eqnarray}
\lefteqn{p_{\chi^2_\nu}(z_1,z_2;\sigma_1,\sigma_2,\tau)}\nonumber\\
&&=\left\{
\begin{array}{l}
\displaystyle
\f{1}{\alpha \tilde \sigma^2 \Gamma(\f{\nu}{2})} \Biggl (
\f{z_1 z_2}{\tau \tilde \sigma^2} \Biggr )^{\f{\nu}{4}-\f{1}{2}}
 \exp \Biggl [ - \f{\sigma_1^{-1}z_1
+\sigma_2^{-1}z_2}
{\alpha \sqrt{2/\nu}} 
           \Biggr ] 
\\
\displaystyle
\quad\times
\textrm{I}_{\f{\nu}{2}-1}
\Biggl( \f{\sqrt{4 \tau z_1 z_2}}{\alpha \tilde \sigma}
\Biggr),
\hfil\textrm{for}\quad 0<\tau<1,
\\
\displaystyle
\f{1}{\alpha \tilde \sigma^2 \Gamma(\f{\nu}{2})} \Biggl (
\f{-z_1 z_2}{\tau \tilde \sigma^2} \Biggr )^{\f{\nu}{4}-\f{1}{2}}
 \exp \Biggl [ - \f{\sigma_1^{-1}z_1
+\sigma_2^{-1}z_2}
{\alpha \sqrt{2/\nu}} 
           \Biggr ] 

\\
\displaystyle
\quad\times\textrm{J}_{\f{\nu}{2}-1}
\Biggl( \f{\sqrt{-4 \tau z_1 z_2}}{\alpha \tilde \sigma}
\Biggr),
\hfil\textrm{for}\quad -1<\tau<0,
\\
\displaystyle
\f{1}{\tilde \sigma^2 (\Gamma(\f{\nu}{2}))^2} \Biggl (
\f{z_1 z_2}{\tilde \sigma^2} \Biggr )^{\f{\nu}{2}-1}
 \exp \Biggl [ - \f{\sigma_1^{-1}z_1
+\sigma_2^{-1}z_2}
{\sqrt{2/\nu}} 
           \Biggr ],
\quad\textrm{for}\quad \tau=0,
\end{array}
\right.
\end{eqnarray}
where $\alpha\equiv 1-\tau$, $\tilde \sigma \equiv \sqrt{2 \sigma_1
\sigma_2 /\nu}$ and $J_n$ and $I_n$ are the Bessel and modified Bessel
function of the first kind, respectively.  The one-point and the
two-point $\chi^2_\nu$ PDFs are scale-invariant and extend from $0$ to
$\infty$ for each variable.  Because we assume that the PDFs of a
density fluctuation have a vanishing mean, we will use off-centered
PDFs, $p^*_{\chi^2_\nu}(z)\!\equiv\!  p_{\chi^2_\nu}(z+\langle z
\rangle)$, $ p^*_{\chi^2_\nu}(z_1,z_2)\!\equiv\!
p_{\chi^2_\nu}(z_1+\langle z_1 \rangle, z_2+\langle z_2 \rangle)$ which
are also scale-invariant in the following analysis.

It is cumbersome to calculate the three-point PDF explicitly.
For simplicity, we only consider the case of $\nu=1$ 
(see also Sheth 1995).
In terms of the covariance matrix for the original trivariate Gaussian PDF,
\BE
\textrm{cov}=
\left(
\begin{array}{@{\,}ccc@{\,}}
s_{11} & s_{12} & s_{13} \\
s_{21} & s_{22} & s_{23} \\
s_{31} & s_{32} & s_{33}
\end{array}
\right ),
\EE
the explicit form of the trivariate $\chi^2_1$ PDF can be written as
\BEA
\nonumber
&& p_{\chi_1^2}(z_1,z_2,z_3;\sigma_1,\sigma_2,\sigma_3,\sigma_{12},
\sigma_{23},\sigma_{13})
\\
\nonumber
&=&  
\f{1}{2 ( 2 \pi)^{3/2} \sqrt{\textrm{det}(\textrm{cov}) z_1 z_2 z_3}}
\exp \Bigl [- \f{ c_{11} z_1+ c_{22} z_2 + c_{33} z_3}
{2} \Bigr]
\\
\nonumber
&\times& 
\Bigl ( \exp(-c_{12}\sqrt{z_1 z_2}) \cosh(c_{23}\sqrt{z_2 z_3}
+c_{13}\sqrt{z_1 z_3})
\nonumber
\\
&+&\exp(c_{12}\sqrt{z_1 z_2}) \cosh(-c_{23}\sqrt{z_2 z_3}
+c_{13}\sqrt{z_1 z_3})
\Bigr ), \label{eq:PDF3}
\EEA
where $\sigma_{ij}=2 s_{ij}^2$ for $i \ne j$, $\sigma_{i}^2=2
s_{ii}^2$, and $c_{ij}=(\textrm{cov}^{-1})_{ij}$.  

It should be noted that the models considered in our analysis
are not exactly identical to the ``$\chi^2_\nu$ field models'' 
(Peebles 1999; Scoccimarro 2000) in which the initial fluctuation itself is 
described by a $\chi^2_\nu$ field.  In contrast to
the Gaussian models, the PDF of a smoothed $\chi^2$ field 
deviates from the original PDF in general (Avelino \& Viana 2000).  If
one would like to explore more realistic models, one should take into
account the dependence of the PDF on smoothing scale. Fortunately, for
$\chi^2_\nu$ PDFs with $\nu$ larger than $10$ smoothed by the Gaussian
filter, it is known that the departure from the original PDF is not
significant although the departure is noticeable in the case of the
sharp $k$-space filter (Avelino \& Viana 2000). Hence, we can expect 
a similar result for some particular choice of smoothing filter.

\section{Solving the cloud-in-cloud problem}
In order to study the characteristics of $\chi^2_\nu$ models on mass
function, we first estimate the conditional probability
\begin{equation}
P(M_1|M_2)\equiv \int_{\delta_c}^\infty  
p^*_{\chi^2_\nu}(\delta_1
\vert \delta_c ;\sigma_1,\sigma_2,\tau)d \delta_1
\end{equation}
for various kinds of smoothing filters, such as the sharp $k$-space, the
Gaussian and the top-hat filters.   

First of all, we consider fluctuations smoothed by the sharp $k$-space
filter $\tilde W_R(k)= 6 R^3 \theta (\pi/R-k)/\pi$. The relation between
the mass $M$ and the smoothing scale $R$ is given by $M=6 R^3 \bar
\rho/\pi$.  The correlation coefficient is 
$\tau=\sigma_2/\sigma_1=(M_1/M_2)^{(n+3)/6}$ (see appendix) provided
that the power spectrum of the initial density fluctuations has a form
${\cal{P}}(k)\propto k^n$, where $n$ denotes the spectral index and
$n>-3$.  The critical value $\delta_c$ is 1.69 independent of mass scale
for the spherical collapse in the Einstein-de Sitter universe. For
simplicity, we assume $\delta_c=1.69$ in the following analysis.  As we
have argued in \S 2, for $M_1 \ll M_2$ and $\sigma_1 \gg \delta_c$, the
EPS approximation gives a correct value,
\begin{eqnarray}
f^{-1}(\textrm{EPS}) 
&=&
\int_0^\infty
p^*_{\chi^2_\nu}(\delta_1;\sigma_1) d \delta_1
\nonumber
\\
&=& \f{\Gamma(\nu/2,\nu/2)}{\Gamma(\nu/2)}\nonumber\\
&=&P(M_{1}|M_{2})\quad \mbox{for}~M_{2}\gg M_{1}~\mbox{and}~\sigma_{1}\gg\delta_{c},
\end{eqnarray}
where $\Gamma(x,y)$ denotes the incomplete Gamma function.  For other
parameter regions, however, the EPS approximation is not always correct, especially
for strongly non-Gaussian PDFs.  From Figure 1, one can see that the
inverse of the EPS factor, $f^{-1}$, deviates from the correct
conditional probability $P(M_{1}|M_{2})$ for a region $M_1 \ll M_2$ and
$\sigma_1 \lesssim \delta_c$, or equivalently $M_1 \gtrsim M_*$, where
$\sigma(M_*)=1$.  For fluctuations smoothed by the sharp $k$-space
filter, the trajectories are described by a Markovian random
walk. Therefore, the chance of upcrossing at the critical value
$\delta_c$ is almost equivalent to the chance of downcrossing at
$\delta_c$, \ti{i.e.}  $P(M_{1}|M_{2})=1/2$.  Therefore, in the
neighborhood of diagonal line $M_1=M_2$, we have $P(M_{1}|M_{2}) \sim
1/2$.  On the other hand, in the Gaussian limit $\nu \rightarrow
\infty$, we have $f^{-1}(\textrm{EPS})=
\Gamma(\nu/2,\nu/2)/\Gamma(\nu/2)\rightarrow 1/2$. Consequently, for
weakly non-Gaussian PDFs smoothed by the sharp $k$-space filter such
that $f^{-1}(\textrm{EPS})\sim 1/2$, it is natural to expect
$P(M_{1}|M_{2})\sim 1/2$ on all scales $M_1 \le M_2$. Thus, for weakly
non-Gaussian PDFs smoothed by the sharp $k$-space filter, the EPS
formalism gives a good approximation of $f$.

Next, we consider the property of fluctuations smoothed by other filters for which the
trajectories of density fluctuations cannot be described by a Markovian
random walk because of the Fourier-mode correlation.
In this case, equation
(\ref{eq:markov}) does not give an exact result.  Let us consider a
trajectory that firstly crosses $\delta_c$ upwards at $M_2$ and ends at
$\delta_{1}>\delta_c$ and $M_1$. For $M_2 \gg M_1$, the trajectory can be
well approximated by that of a Markovian random walk, since the
correlation length is almost negligible compared with the length of the
whole trajectory.  On the other hand, for $M_2 \sim M_1$, the trajectory
can be well approximated by a monotonically increasing function in
decreasing value of $M$, or increasing value of $\sigma$, since the correlation
coefficient is almost equal to unity.  
Then the conditional probability is approximately given by
\begin{equation}
P(M_1 \vert M_2)
\approx 
\int_{\delta_c}^\infty  
p(\delta_1 \vert \delta_2=\delta_c, \del \delta_2 / \del M_2<0)
d \delta_1. \label{eq:conditionN}
\end{equation}
At large scales, $\delta_1 \ll \delta_c$, or equivalently $M_1 \gg M_*$,
the probability of upcrossing at $\delta_c$ is very low. In other words,
the probability of $\del \delta_2 / \del M_2 \ge 0$ at
$\delta_2=\delta_c$ is almost zero.  Therefore, the extra condition
$\del \delta_2 / \del M_2>0$ is not necessary at large scales.  As shown
in Figure 2, when the Gaussian filter is used, $P(M_1 \vert M_2)$
increases compared with that for the sharp $k$-space filter owing to the
Fourier-mode correlation.  At large scales $M_1 \gg M_*$, $P(M_1\vert
M_2)\sim 1$. Consequently, the chance of downcrossing at $\delta_c$ is
almost zero.  Thus it is reasonable to conclude that equation
(\ref{eq:markov}) still provides a good approximation of mass functions,
at least, at large scales even for other filters.  On the other hand, in
the region $M_{1}\ll M_{*}$ and $M_{1}\ll M_{2}$, $P(M_{1}|M_{2})$ is
close to the value in the case of the sharp $k$-space filter.  This
suggests that the approximation of the Markovian random walk is good in
that parameter region.

Substituting the variances $\sigma_1$, $\sigma_2$, and the correlation
coefficient $\tau$ for each type of smoothing filter (see appendix for
derivation) into the conditional probability equation (\ref{eq:markov}),
one can compute the mass function by solving the integral equation
(\ref{eq:jed0}).  In Figure 3, we show the multiplicity functions $F(M)$
for the various filters and for the EPS formalism in the case of
$\chi^{2}_{1}$ PDF with one degree of freedom.  The multiplicity
functions derived here decrease at large scales compared with those of
the EPS prediction.  As we have argued in \S 2, the actual value of
$P(M_{1}|M_{2})=p(\delta_1 \ge \delta_c \vert \delta_2=\delta_c)$ is
larger than the value $f^{-1}$ predicted by using the EPS approximation
at large scales, which explains a decrease in the multiplicity function
$F(M)$.  To compensate the deficit at large scales, $F(M)$ increases at
small scales.  Regarding the dependence of $F(M)$ on the spectral index
$n$, one can see in Figure 3 that a relative increase of $F(M)$ at small
scales is much prominent for larger $n$. This is because $F(M)$
decreases rapidly at smaller scales for a larger value of $n$, or for a
much \ti{blue} spectrum.  Similar behavior is also observed for the
Gaussian models (Nagashima 2001).

In Figure 4, we show, for $\chi^2_\nu$ PDFs with various degree of
freedom, $dg(\sigma)/d \sigma$ which is defined as $g(\sigma)=\tilde g
(M(\sigma))$ for which $\tilde g (M)=\int_M^\infty F(M') dM'$ is the
fraction of collapsed objects above a smoothing scale $M$. At large
scales, $\sigma \lesssim 1$, $dg(\sigma)/d \sigma$ is significantly
increased compared with those corresponding to the Gaussian PS mass
function.  This is because the $\chi^2_\nu$ PDFs have a broad tail
toward large $\delta$, especially for strongly non-Gaussian PDFs with
small value of $\nu$. Even in the case of $\nu=50$, one can still
observe a clear difference from the value corresponding to the PS mass
function for fluctuations with small variance $\sigma \lesssim 1$.  In
other words, the amount of rare density peaks at large scales is very
sensitive to the non-Gaussianity of the initial fluctuations.  In the
Gaussian limit, $\nu \rightarrow \infty$, the EPS formalism correctly
reproduces $dg(\sigma)/d \sigma$ for fluctuations smoothed by the sharp
$k$-space filter. As the degree of freedom $\nu$ decreases, a deviation
from the EPS prediction becomes noticeable. It is clear that the EPS
formalism overestimates the number density of dark halos on large scales
$\sigma \lesssim 1$, especially for strongly non-Gaussian PDFs.  For
fluctuations smoothed by the Gaussian and the top-hat filters, such a
deviation is prominent even in the case of a weakly non-Gaussian PDF
($\nu=50$) owing to the correlation between Fourier-modes.  Similar
result has been obtained by Avelino \& Viana (2000) using Monte Carlo
simulations for smoothed $\chi^2_\nu$ fields.

Note that our analytic calculation of 
$dg(\sigma)/d \sigma$ for $\nu=10$ and $\sigma \lesssim
1$ agrees well with the previous 
numerical results by Avelino \& Viana (2000) when
the Gaussian filter is used.  The result is natural, since  
the departure from the original one-point PDF is not 
significant in this case. However, for fluctuations smoothed by the Gaussian
filter with slightly large variance $1 \!\lesssim\!\sigma\!\lesssim\!5$,
our analytic values exceed their numerical values. We might be able to
recover their numerical results if we use the improved approximation
described by equation (\ref{eq:conditionN}) of the conditional
probability instead.  Nevertheless, for the purpose of constraining
non-Gaussianity of initial fluctuations using the abundance of
high-redshift clusters (Matarrese, Verde \& Jimenez 2000), our
approximation of the conditional probability may be sufficient in
practice, as mentioned.

\section{Effect of mode correlation }
So far, we have assumed that different Fourier-modes of $\chi^2_\nu$ 
fluctuations are uncorrelated. However, 
if we consider $\chi^2_\nu$ field models
defined as $\phi(\textbf{x})\equiv \sum_i^{\nu} 
\psi_i^2(\textbf{x})$ where $\psi_i(\textbf{x})$ 
are independent homogeneous and isotropic random Gaussian 
fields whose Fourier-modes are uncorrelated, 
Fourier-modes $\phi_{\textbf{k}}$ of a
$\chi^2_\nu$ field are no longer uncorrelated. 
The signature of correlation appears as non-vanishing skewness
, kurtosis, or higher-order cumulants  
although the two-point correlation has the same form as 
the Gaussian one (Scoccimarro 2000). 
In this case, a hypothetical 
particle with position $\delta(M)$ cannot be described by 
the Markovian random walk even if we smooth the fluctuations by 
using the sharp-k space filter. Therefore, it is of 
crucial importance to estimate the
effect of correlation.

In order to check the 
validity of our calculation, we compare the 
conditional probability $P_3\equiv p(\delta_1 \ge \delta_c |
\delta_2=\delta_c, \delta_3 < \delta_2),M_1 \le M_2 \le M_3$ 
specified by a three-point
and two-point PDFs to 
$P_2 \equiv p(\delta_1 \ge \delta_c |\delta_2=\delta_c)$ specified 
by a two-point and one-point PDFs. In fact, the actual condition 
to be satisfied is equation (\ref{eq:P12}), which is equivalent to 
$P_n, n\rightarrow \infty$. $P_3$ is expected to be more accurate  
than $P_2$ for estimating the conditional probability $P(M_1 \vert M_2)$.

In what follows, we only consider $\nu=1$ case, 
for which the effect of mode correlation owing to the non-Gaussianity 
is most significant. In the limit $\sigma_3 \rightarrow 0$, the
correlation coefficients vanish
$\tau_{i3}\sim \sigma_3/\sigma_i \rightarrow 0$, which gives 
$p(\delta_1 \vert \delta_2, \delta_3)=
p(\delta_1,\delta_2)p(\delta_3)/p(\delta_2)p(\delta_3)=
p(\delta_1 \vert \delta_2)$. Thus, if $M_3$ is large enough compared with 
$M_2$ and $M_1$, the effect of mode correlation is negligible.
On the other hand, for $M_3 \sim M_2$, or equivalently, 
$\sigma_3 \sim \sigma_2$, the probability of
having a trajectory $\delta_1 \ge \delta_c$ becomes large
owing to the correlation in the fluctuations on different scales.
Note that in the limit
$\sigma_3 \rightarrow \sigma_2$, $P_3$ is equivalent to the
right-hand-side of equation (\ref{eq:conditionN}) assuming that 
$\delta_2(\sigma_2)$ is differentiable at $\sigma_2$.
From equation (\ref{eq:PDF3}), one can easily calculate the ratio $P_3/P_2$. 

First of all, we consider the case of the Gaussian filtering.  
It turns out that the ratio $P_3/P_2$ is $\lesssim 1.5$ for $\sigma_1=2$, and
$\lesssim 1.2$ for $\sigma_1=0.5$ if the spectral index is $n=-2$
(see figure 5). At large scales $\sigma < 1$, a decrease in the 
multiplicity function $F(M)$ is expected to be less than 
$\sim 30$ percent.     

In the case of the sharp-k space filtering, we have the scale-invariant 
relations in the covariances: $\sigma_{12}=\sigma_2^2, \sigma_{23}
=\sigma_{3}^2, \sigma_{13}=\sigma_{3}^2$. Then we obtain
\BEA
&&
p_{\chi_1^2}^{SK}(z_1 |z_2,z_3)
\\
&=&
p_{\chi_1^2}^{SK}(z_1 |z_2)
\nonumber
\\
&=& 
\f{1}{\sqrt{2 \pi z_1 (s_1^2-s_2^2)}}\exp \Biggl ( -\f{z_1+z_2}
{2(s_1^2-s_2^2)}\Biggr)
\cosh \Biggl ( \f{\sqrt{z_1 z_2}}{s_1^2-s_2^2}
\Biggr),
\EEA
where $\sigma_i^2=2 s_i^4$, which gives $P_3=P_2$ irrespective of 
the value of the spectral index $n$ as in the case of the Gaussian 
fluctuations. Thus, the 
effect of three-point correlation is completely 
negligible if the fluctuations are
smoothed by using the sharp-k space filter.  
This relation might also hold for
other $\chi^2_\nu$ PDFs with $\nu>1$, since the effect of 
correlation is less significant. Presumably, all the scale-invariant
PDFs have this property. 

To summarize, the effect of mode correlation is less significant  
for fluctuations at large scales $\sigma < 1 $ 
irrespective of the type of filtering. 
Inclusion of an additional condition $\delta_3 < \delta_c$
improves the reliance of calculation for strongly non-Gaussian
fluctuations at small scales $\sigma>1$.

\section{Conclusions}
In this paper, we have demonstrated that the EPS approach has a drawback
in describing the number of collapsed objects, especially for strongly
non-Gaussian density fluctuations.  Based on the formalism developed by 
Jedamzik, and Yano, Nagashima and Gouda properly taking into account the scale 
dependence of correlation between objects at different scales, 
we analytically calculated the mass
function for various $\chi^2_\nu$ models and found a deviation from
those predicted by using the EPS formalism, especially noticeable for
strongly non-Gaussian models: a decrease at large scales and an increase
at small scales in the value of multiplicity function.  The result may
affect some recent studies of constraints on non-Gaussian models using
the cluster abundance at different red shifts and the correlation length
of galaxy clusters (Chin, Ostriker \& Strauss 1998; Koyama, Soda \&
Taruya 1999; Robinson, Gawiser \& Silk 2000). Our results are similar to 
those by Avelino and Viana (2000) based on Monte-Carlo
simulations of non-Gaussian $\chi^2_\nu$ fields.  At intermediate
scales, $1 \lesssim \sigma \lesssim 5 $, the deviation from the EPS prediction is
not prominent. It seems that the result is consistent with those
from $N$-body simulations for various non-Gaussian fields (Robinson \&
Baker 2000).  It would be interesting if larger $N$-body simulation could
be carried out and find out a deviation of mass function from the EPS
prediction for objects with very large mass (low-$\sigma$) or for those
with very small mass (high-$\sigma$).

In order to vindicate that the 
generalized PS formalism works in estimating the mass function
for non-Gaussian models, we should take various kinds of effects into
consideration: effects of non-spherical collapse (for Gaussian fields,
see Sheth et al 2001), ambiguity in
mass-smoothing scale relation (Bardeen et al 1986, Peacock \& 
Heavens 1991), and conditions of objects surrounding
by an isolated dark halo. The last issue is relevant to the
cloud-in-cloud problem.  Although we have discussed about a prescription
for incorporating the condition of upcrossing at the critical value
$\delta_c$, we did not explicitly consider the effect of spatial
correlation owing to the finite size of halos in evaluating the
conditional probability $P(M_1 \vert M_2)$. For Gaussian models it is
known that the effect of spatial correlation almost cancels out the
filtering effect, recovering the original PS mass function, particularly
in the case of the top-hat filter (Nagashima 2001). It is of very
importance to check whether such a cancellation occurs for non-Gaussian
models.  The other issues left untouched should be addressed
in our future work.

\acknowledgments We would like to thank N. Seto and A. Taruya for useful
discussion and comments. We would also like to thank the anonymous
referee for useful suggestion and comments.

\appendix 

\section{Correlation coefficients}

In this section, we derive the formulae of correlation coefficients
$\tau$ for density fluctuations smoothed by three types of filter,
namely, the sharp $k$-space filter, the Gaussian filter and the top-hat
filter.

Let us consider a mass density fluctuation (contrast) $\delta ({\x})
\equiv \frac{\rho ({\x})-\bar\rho}{\bar\rho}$, where $\rho({\x})$
denotes the mass density at a point ${\x}$ and $\bar\rho$ is the mean
cosmic mass density.  Using a window function $W_R(r=|x|)$, $\delta
({\x})$ can be smoothed on scale $R$
\begin{equation}
\delta_R(\x)=\int  W_R(\vert \x-\x' \vert)
\delta(\x')d{\bf x},
\end{equation}
and in Fourier space,
\begin{equation}
~~=\f{1}{(2 \pi)^3} \int  \tilde W_R(k)
\delta(\K)e^{i{\bf k\cdot x}}d{\K},
\end{equation}
where $\delta_\K$ and $\tilde W_R(k)$ are the Fourier transforms of $\delta_\K$
and $W_R(r)$, respectively.  Here we choose a normalization of the
window function as $W_R(0)=1$.  
\\
\indent
If the fluctuations are homogeneous,  then the two-point 
correlation is written in terms of the
power spectrum ${\cal{P}}(k)$ as
\begin{equation}
\langle \delta^{}_{\K_1}\delta^*_{\K_2} \rangle =(2 \pi)^3 
\delta_D (\K_1-\K_2){\cal{P}}(k_1). \label{eq:a3} 
\end{equation}
which gives 
\begin{equation}
\langle 
(\delta_R)^2 \rangle =\f{1}{( 2 \pi)^3} \int_0^\infty 4
\pi k^2 \tilde W_R^2(k) {\cal{P}}(k) dk. \label{eq:a4}
\end{equation}
The correlation coefficient is given by $\tau\equiv 
\langle \delta_{R_1} \delta_{R_2}\rangle
/\sigma_{R_1} \sigma_{R_2}$ 
where
\begin{equation}
\langle \delta_{R_1} \delta_{R_2} \rangle 
= \f{1}{( 2 \pi)^3}  
\int_0^\infty 4 \pi k^2 \tilde W_{R_1}(k) \tilde W_{R_2}(k) 
{\cal{P}}(k) dk. \label{eq:a5}
\end{equation}
Note that all the equations in this appendix except for (\ref{eq:a3})
hold for fluctuations whose Fourier transforms are correlated.   
The mass of objects smoothed on
scale $R$ can be defined as
\BE
M \equiv \int {\bar \rho} W_R(\x) d \x.
\EE
The correlation coefficients for the
three types of filters 
are written as follows. We assume that ${\cal{P}}(k) \propto k^n$. 

\begin{enumerate}
\item Sharp $k$-space filter\\
\begin{eqnarray}
W_R(r)&=& \f{3 }{k_c^3 r^3}(\sin k_{c}r -k_{c}r\cos k_{c}r),\\
\tilde W_R(k)&=& \f{6 \pi^2}{k_c^3}\theta(k_{c}-k),
\end{eqnarray}

where $k_{c}$ is the cut-off wave number, $k_{c}\simeq R^{-1}$, and
$\theta(x)$ is the Heaviside step function. 
Note that there is an ambiguity in the relation between $k_c$ and $R$. Here we
define it as $k_c=\pi/R$ that gives $M=6 R^3 \bar \rho/\pi $ which has
been widely used in the literature. If we
choose $k_c=(9 \pi/2)^{1/3}/R$, then we have $M=4 \pi R^3 \bar \rho /3$.
For the former definition, the variance of density fluctuation is
\begin{equation}
\sigma_M^2 \propto \f{R^{-3-n} \pi^{1+n}}{2 (3+n)}, 
\end{equation}
and the correlation coefficient for $M_1<M_2$ is
$\tau=\sigma_{M_2}/\sigma_{M_1}=(M_2/M_1)^{-(n+3)/6}=(R_2/R_1)^{-(n+3)/2}$.

\item Gaussian filter\\
\begin{eqnarray}
W_R(r)&=&  \exp(-r^2/(2 R^2)),\\
\tilde W_R(k)&=&(2 \pi)^{3/2}R^3 \exp(-k^2 R^2/2).
\end{eqnarray}
The mass of objects smoothed on
scale $R$ is $M=(2 \pi)^{3/2}R^3 \bar \rho$ and for $n>-3$,
\begin{eqnarray}
\sigma_M^2
&\propto&( 2 \pi)^{-2}R^{-3-n} \Gamma((3+n)/2),
\\
\tau
&=&
\f{2^{(n+3)/2}(\sigma_1^{-4/(3+n)}+\sigma_2^{-4/(3+n)})^{-(3+n)/2}
       }{\sigma_1 \sigma_2}.
\end{eqnarray}

\item Top-hat filter\\
\begin{eqnarray}
W_R(r)&=&\theta (1-r/R),\\
\tilde W_R(k)&=& 4 \pi R^3 \Biggl( \f{\sin kR}{(k R)^3}-
\f{\cos kR}{(k R)^2} \Biggr).
\end{eqnarray}
The mass of objects smoothed on
scale $R$ is
$M=4 \pi R^3 \bar \rho/3$. If $n$ is not an integer, 
for $-3<n<1$,  
\begin{equation}
\sigma_M^2
\propto \f{9}{2 \pi^2 R^3} (2 R)^{-n} (n-2)(n+1)\sin (n \pi/2) \Gamma(-3+n),
\end{equation}

\begin{eqnarray}
\nonumber
\tau
&=&\f{9 \sin(n\pi/2)}{4 \pi^2 R_1^3 R_2^3}
\Biggl(\biggl( (R_1+R_2)^{3-n}-(R_2-R_1)^{3-n}
\biggr)(\Gamma(-3+n)+\Gamma(-2+n))
\\
&+& \biggl( (R_1+R_2)^{1-n}-(R_2-R_1)^{1-n}
\biggr)R_1 R_2 \Gamma(-1+n)
\Biggr)\sigma_1^{-1} \sigma_2^{-2}.
\end{eqnarray}
For integers $-3<n<1$,
\begin{equation}
\sigma_M^2 \propto \f{9}{2 \pi^2 R^3}(2 R)^{-n}\times
\left\{
\begin{array}{ll}
\displaystyle{
\f{\pi}{6}}~~~(n=0)&\\
\\
\displaystyle{\f{1}{8}~~~(n=-1) }&\\
\\
\displaystyle{\f{\pi}{60}~~~(n=-2)}&,
\end{array}\right.\\
\end{equation}
and
\begin{equation}
\tau=
\left\{
\begin{array}{ll}
\displaystyle{
\sqrt{\f{R_1^3}{R_2^3}}=\f{\sigma_2}{\sigma_1},~~(n=0)}&\\
\\
\displaystyle{\f{1}{4 R_1^2 R_2^2}\biggl (2R_1 R_2
 (R_1^2+R_2^2)+(R_2^2-R_1^2)^2
\ln\f{R_1-R_2}{R_1+R_2} \biggr),~~~(n=-1) }&
\\
\\
\displaystyle{\f{5 R_2^2-R_1^2}{4 R_2^3}\sqrt{R_1 R_2}
=\f{\sigma_2}{4 \sigma_1^5}(5\sigma_1^4-\sigma_2^4),~~~(n=-2)}.&
\end{array}\right.\\
\end{equation}
\end{enumerate}

\begin{figure*}
\centerline{\includegraphics[width=13cm]{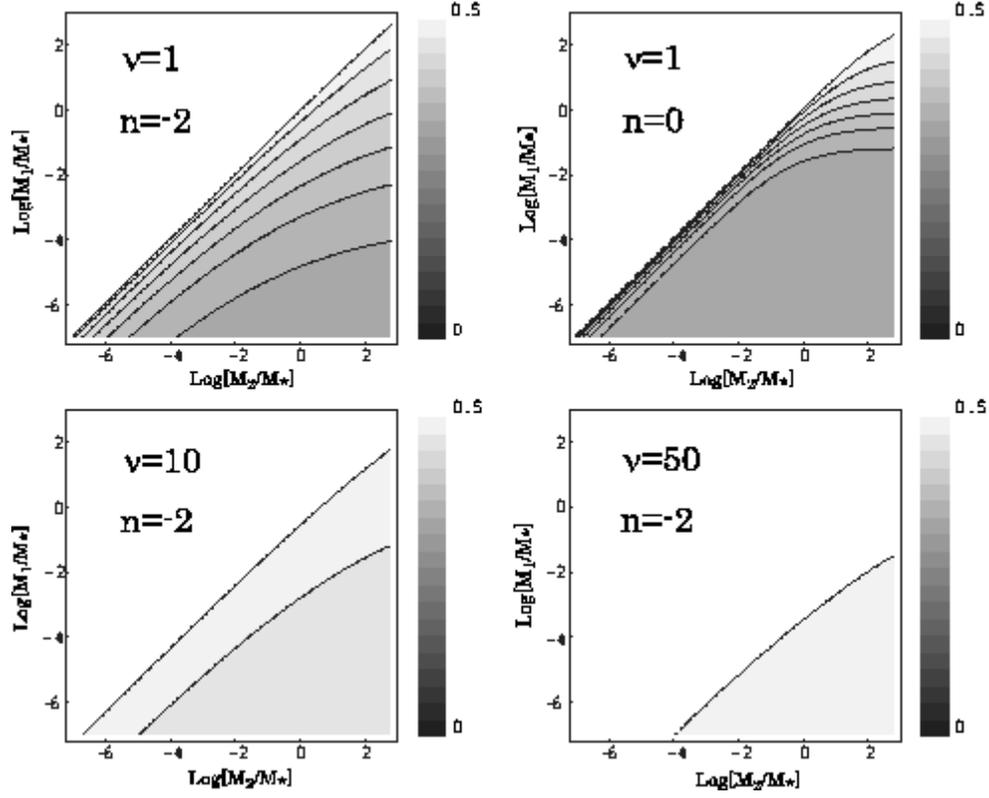}}
\caption{Plots of the conditional probability $p(\delta_1 \ge \delta_c \vert
\delta_2=\delta_c;\sigma(M_1),\sigma(M_2),\tau(M_1,M_2))$ for
$\chi^2_\nu$ PDFs smoothed by the sharp $k$-space filter. $n$ denotes the
spectral index. The accompanying palettes show the relation between the
gray level and the value. $M_*$ is defined as $\sigma(M_*)=1$. We set 
$\delta_c=1.69$.   
}
\end{figure*}

\begin{figure*}
\centerline{\includegraphics[width=13cm]{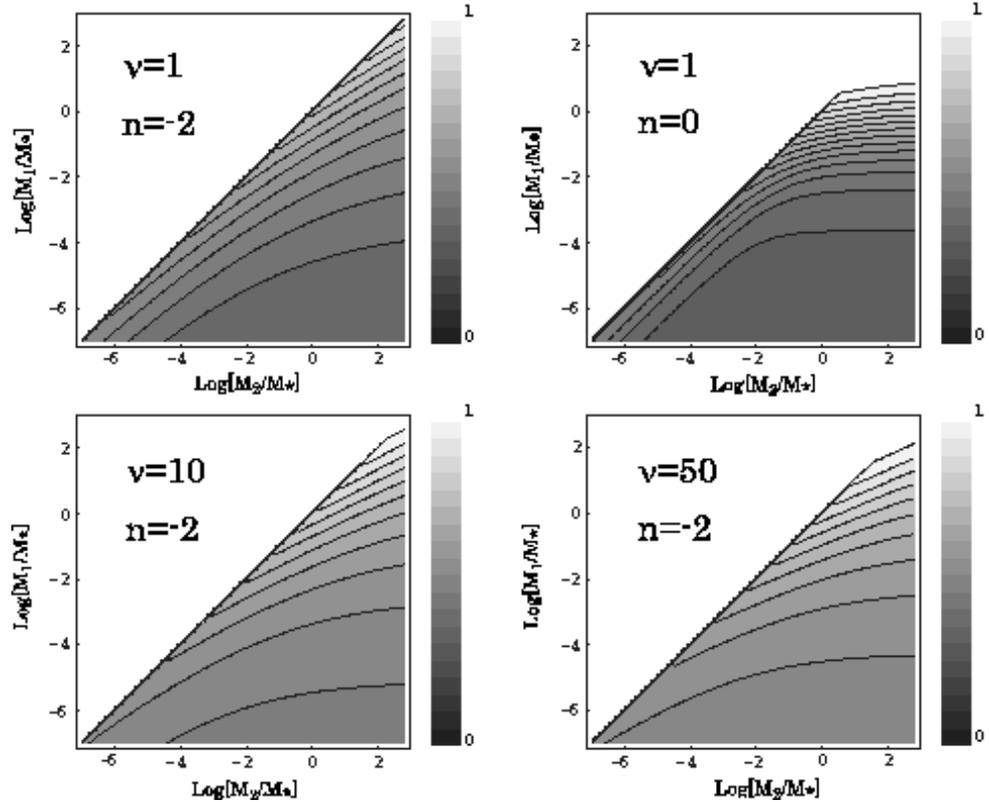}}
\caption{Same as Figure 1, but for the Gaussian filter.  Note that the
 contour level is from 0 to 1.}
\end{figure*}

\begin{figure*}
\epsscale{0.8}

\centerline{\includegraphics[width=9cm]{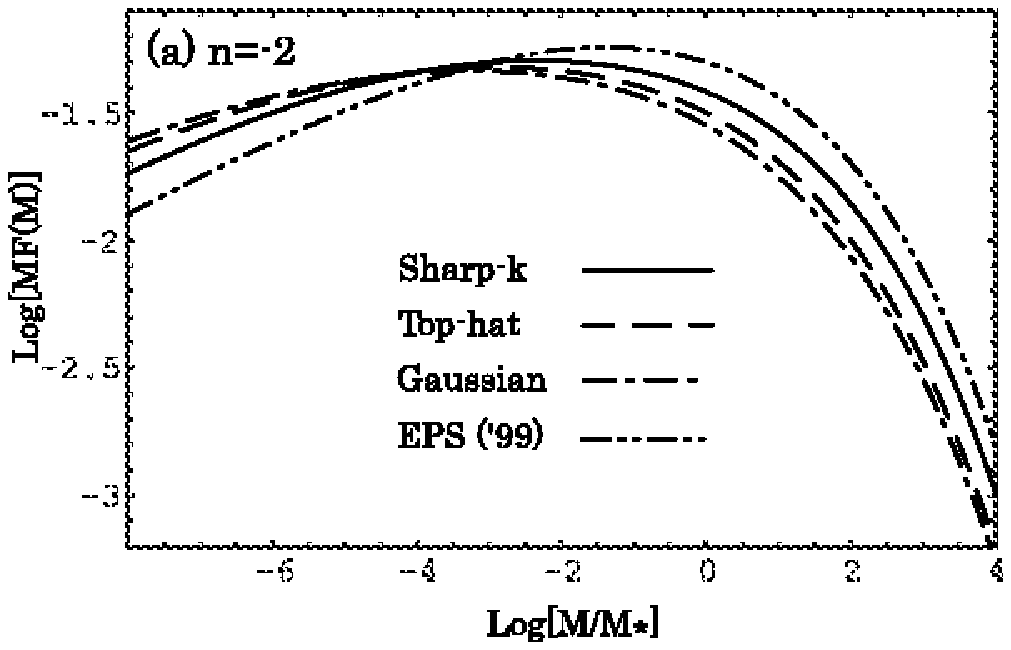}}
\centerline{\includegraphics[width=9cm]{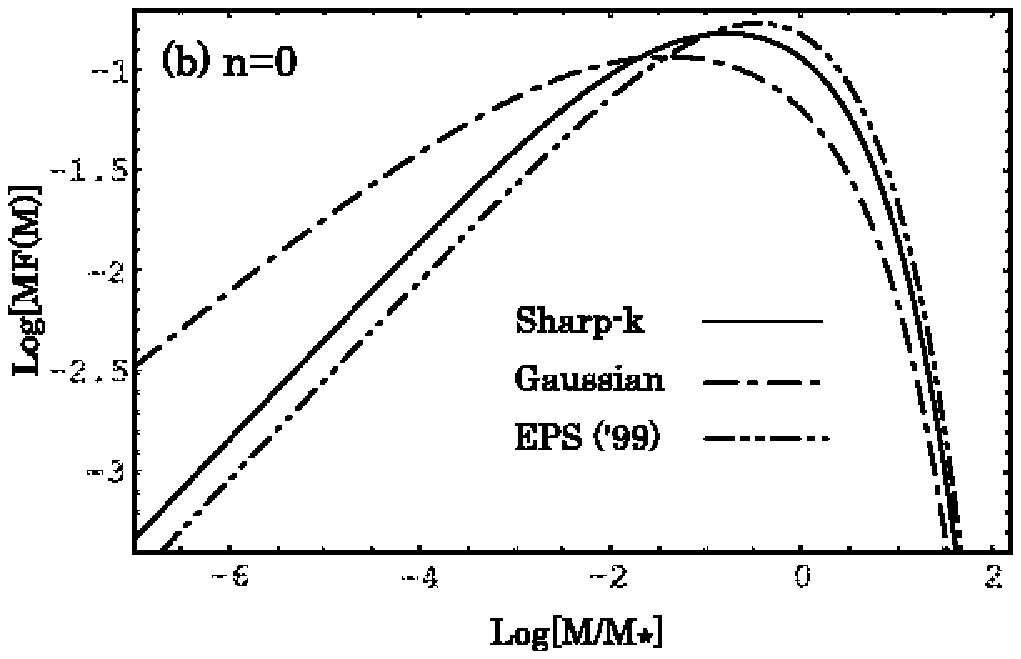}}

\caption{Plots of multiplicity functions $F(M)=M n(M)/\bar \rho$ for
fluctuations obeying the $\chi^2_1$ PDF smoothed by various
filters. (a)$n=-2$. (b)$n=0$. The solid, the dashed, and the
dashed-dotted lines correspond to $F(M)$ for fluctuations smoothed by
the sharp $k$-space filter, the top-hat filter, and the Gaussian filter,
respectively. The dashed-double-dotted line represents the value
predicted by using the EPS formalism. $M_*$ is defined as
$\sigma(M_*)=1$.  In the case of $n=0$, the multiplicity function $F(M)$
for fluctuations smoothed by the top-hat filer is exactly the same as
that smoothed by the sharp $k$-space filter. }

\end{figure*}

\begin{figure*}
\centerline{\includegraphics[width=9cm]{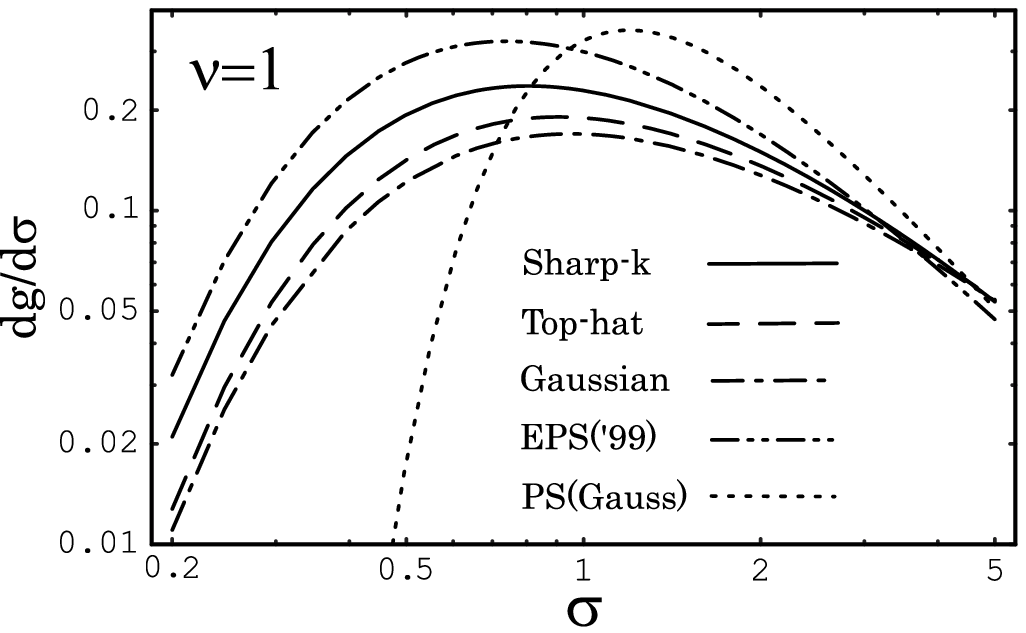}}
\centerline{\includegraphics[width=9cm]{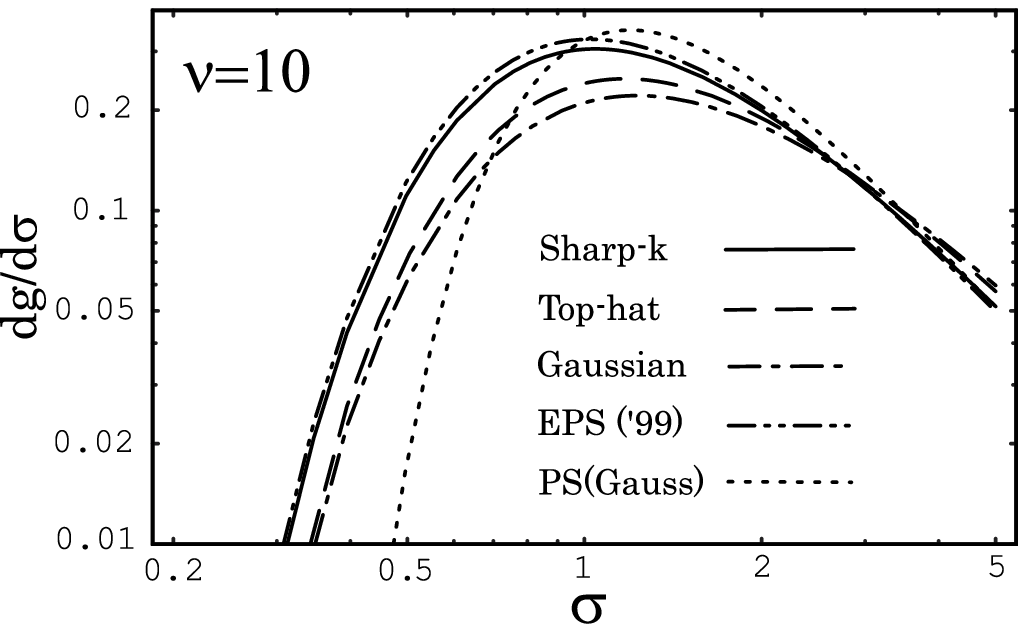}}
\centerline{\includegraphics[width=9cm]{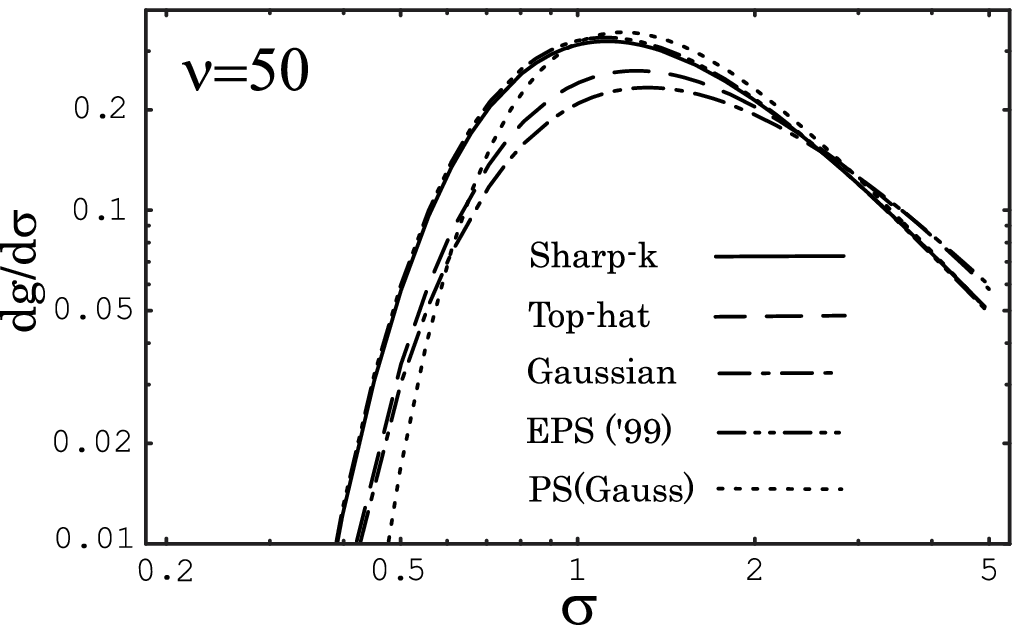}}

\caption{Plots of $d g /d \sigma$ for fluctuations that obey three kinds
of $\chi^2$ PDFs ($\nu=1, 10$ and 50).  The spectral index is assumed to
be $n=-2$.  The solid, the dashed, and the dashed-dotted lines
correspond to $F(M)$ for fluctuations smoothed by the sharp $k$-space
filter, the top-hat filter and the Gaussian filter, respectively. The
dashed-double-dotted line represents the value predicted by the EPS
formalism. The dotted line corresponds to the value predicted by the PS
formalism for the Gaussian PDF. In the case of the sharp-k space filter,
$d g /d \sigma$ does not depend on $n$.  }

\end{figure*}
\begin{figure*}

\centerline{\includegraphics[width=9cm]{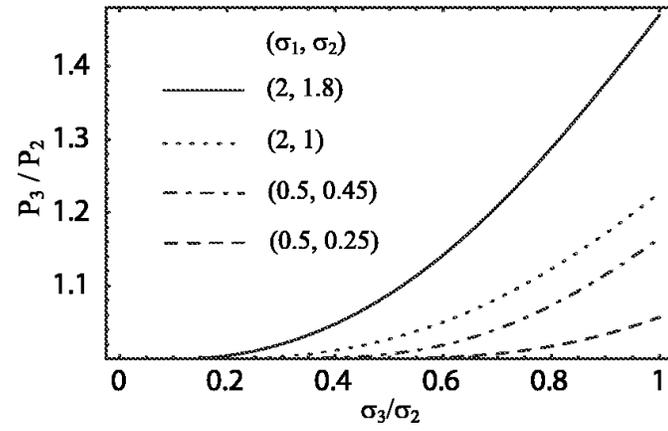}}
\caption{Effect of mode correlation in the case of Gaussian filtering
 for $\chi^2_1$ fluctuations with the spectral index $n=-2$. A 
deviation from 1 in the value of $P_3/P_2$ gives a relative error
 in $P_2$ for estimating the conditional probability $P(M_1 \vert M_2)$
 assuming that the effects of four-point or higher-order 
correlations are negligible. }

\end{figure*}

\end{document}